\documentclass[10pt,conference]{IEEEtran}
\IEEEoverridecommandlockouts

\usepackage{algorithm}
\usepackage{algpseudocode}
\usepackage{standalone}
\usepackage{graphicx}
\usepackage{multicol}
\usepackage{skak}
\usepackage{tlatex}
\usepackage{mathtools}
\usepackage{enumerate}
\usepackage{verbatim}
\usepackage{url}
\usepackage[]{soul}
\usepackage[textsize=scriptsize]{todonotes}

\usepackage{pifont}
\newcommand{\xmark}{\ding{55}}%

\usepackage{amsmath}
\usepackage{amssymb}

\usepackage{tikz}
\usetikzlibrary{positioning,fit,calc}
\usetikzlibrary{shapes.geometric}

\newcommand{\tlaplus}{TLA\textsuperscript{+}}

\newif\ifanonymous

\begin{document}

\title{Plain and Simple Inductive Invariant Inference for Distributed Protocols in \tlaplus
\thanks{This work was supported by the U.S. National Science Foundation under NSF SaTC award CNS-1801546.}
}



\author{
\IEEEauthorblockN{William Schultz}
\IEEEauthorblockA{
\textit{Northeastern University}\\
Boston, MA \\
schultz.w@northeastern.edu}
\and  
\IEEEauthorblockN{Ian Dardik}
\IEEEauthorblockA{
\textit{Carnegie Mellon University}\\
Pittsburgh, PA \\
idardik@andrew.cmu.edu}
\and
\IEEEauthorblockN{Stavros Tripakis}
\IEEEauthorblockA{
\textit{Northeastern University}\\
Boston, MA \\
stavros@northeastern.edu}
}

\newcommand{\mytodo}[1]{\noindent{\textcolor{red}{TODO: #1}}}
\newcommand{\todowill}[1]{\todo{#1 (Will)}}
\newcommand{\toolendive}{\textit{endive}}

\newcommand{\red}[1]{{{\textcolor{red}{#1}}}}
\newcommand{\green}[1]{{{\textcolor{green}{#1}}}}
\newcommand{\blue}[1]{{{\textcolor{blue}{#1}}}}
\newcommand{\gray}[1]{{{\textcolor{gray}{#1}}}}
\newcommand{\orange}[1]{{{\textcolor{orange}{#1}}}}

\newcommand{\ctis}{{CTIs}}
\newcommand{\invs}{{Invs}}

\newcommand{\sts}{STS\,}
\newcommand{\safe}{Safe}

\newcommand{\nlemmasparam}{N_{lemmas}}
\newcommand{\numbenchmarkssolved}{25}
\newcommand{\numbenchmarks}{29}

\renewcommand{\implies}{\Rightarrow}

\setstcolor{red}

\newcommand{\add}[1]{#1}
\newcommand{\remove}[1]{}

\maketitle



\begin{abstract}
    We present a new technique for automatically inferring inductive invariants of parameterized distributed protocols specified in \tlaplus. Ours is the first such invariant inference technique to work directly on \tlaplus, an expressive, high level specification language. To achieve this, we present a new algorithm for invariant inference that is based around
    %
    a core procedure for generating \textit{plain}, potentially non-inductive \textit{lemma invariants} that are used as candidate conjuncts of an overall inductive invariant. 
    We couple this with a greedy lemma invariant selection procedure that selects lemmas that eliminate the largest number of counterexamples to induction at each round of our inference procedure.
    We have implemented our algorithm in a tool, \toolendive{}, and evaluate it on a diverse set of distributed protocol benchmarks,
    demonstrating competitive performance and ability to uniquely solve an industrial scale reconfiguration protocol.
\end{abstract}

\section{Introduction}

Automatically verifying the safety of distributed systems remains an important and difficult challenge. Distributed protocols such as Paxos \cite{lamport1998the} and Raft \cite{raftpaper} serve as the foundation of modern fault tolerant systems, making the correctness of these protocols critical to the reliability of large scale database, cloud computing, and other decentralized systems \cite{2020cockroachdb,2021awsfms3,2020formalspec_tendermint,newcombe2014use}. An effective approach for reasoning about the correctness of these protocols involves specifying system \textit{invariants}, which are assertions that must hold in every reachable system state. Thus, a primary task of verification is proving that a candidate invariant holds in every reachable state of a given system. For adequately small, finite state systems, symbolic or explicit state model checking techniques \cite{burch1992symbolic,holzmann1997model,biereBMC} can be sufficient to automatically prove invariants. For verification of infinite state or \textit{parameterized} protocols, however, model checking techniques may, in general, be incomplete \cite{bloem2015decidability}. Thus, the standard technique for proving that such a system satisfies a given invariant is to discover an \textit{inductive invariant}, which is an invariant that is typically stronger than the desired system invariant, and is preserved by all protocol transitions. Discovering inductive invariants, however, is one of the most challenging aspects of verification and remains a non-trivial task with a large amount of human effort required \cite{woos2016planning,chand2016formal,wilcox2015verdi,schultz2021formal}. Thus, automating the inference of these invariants is a desirable goal.


In general, the problem of inferring inductive invariants for infinite state protocols is undecidable \cite{padon2016decidability}. Even the verification of inductive invariants may require checking the validity of arbitrary first order formulas, which is undecidable \cite{padonpaxosEPR}. Thus, this places fundamental limits on the development of fully general algorithmic techniques for discovering inductive invariants. 



Significant progress towards automation of inductive invariant discovery for infinite state protocols has been made with the Ivy framework \cite{padon2016ivy}.
Ivy utilizes a restricted system modeling language that allows for efficient checking of verification goals via an SMT solver such as Z3 \cite{de2008z3}.
In particular, the EPR and extended EPR subsets of Ivy are decidable.  Ivy also provides an interface for an interactive, counterexample guided invariant discovery process. 
The Ivy language, however, may place an additional burden on users when protocols or their invariants don't fall naturally into one of the decidable fragments of Ivy. Transforming a protocol into such a fragment is a manual and nontrivial task \cite{padonpaxosEPR}. 
%
    
Subsequent work has attempted to fully automate the discovery of inductive invariants for distributed protocols. State of the art tools for inductive invariant inference for distributed protocols include I4 \cite{Ma2019}, fol-ic3 \cite{2020firstorderquantified}, IC3PO \cite{goel2021symmetry}, SWISS \cite{hance2021finding}, and DistAI \cite{yao2021distai}. All of these tools, however, accept only Ivy or an Ivy-like language \cite{mypyvyTool} as input. Moreover, several of these tools work only within the restricted decidable fragments of Ivy.

In this paper, we present a new technique for automatic discovery of inductive invariants for protocols specified in \tlaplus{}, a high level, expressive specification language \cite{lamport2002specifying}. To our knowledge, this is the first inductive invariant discovery tool for distributed protocols in a language other than Ivy. Our technique is built around a core procedure for generating small, \textit{plain} (potentially non-inductive) invariants. We search for these invariants on finite protocol instances, employing the so-called \textit{small scope} hypothesis \cite{softwareabstractions,Ma2019,arons2001parameterized}, circumventing undecidability concerns when reasoning over unbounded domains. We couple this invariant generation procedure with an invariant selection procedure based on a greedy counterexample elimination heuristic in order to incrementally construct an overall inductive invariant.
By restricting our inference reasoning to finite instances, we avoid restrictions imposed by modeling approaches that try to maintain decidability of SMT queries.





Our technique is partially inspired by prior observations \cite{chand2016formal,schultz2021formal,hance2021finding,bradley2011sat} that, for many practical protocols, an inductive invariant $I$ is typically of the form $I = P \wedge A_1 \wedge \dots \wedge A_n$, where $P$ is the main invariant (i.e. safety property) we are trying to establish, and $A_1,\dots,A_n$ are a list of \emph{lemma invariants}. Each lemma invariant $A_i$ may not necessarily be inductive, but it is necessarily an invariant, and it is typically much smaller than $I$. These lemma invariants serve to strengthen $P$ so as to make it inductive. Many prior approaches to inductive invariant inference have focused on searching for lemma invariants that are inductive, or inductive relative to previously discovered information \cite{hance2021finding,bradley2011sat,goel2021symmetry,2020firstorderquantified}. In contrast, our inference procedure searches for \textit{plain} lemma invariants and uses them as candidates for conjuncts of an overall inductive invariant. To search for lemma invariants, we sample candidates using a syntax-guided approach \cite{2018accelsynsyninvs}, and verify the candidates using an off the shelf model checker.

We have implemented our invariant inference procedure in a tool, \toolendive{}, and we evaluate its performance on a set of diverse protocol benchmarks, including \numbenchmarks{} of the benchmarks reported in \cite{goel2021symmetry}. Our tool solves nearly all of these benchmarks, and compares favorably with other state of the art tools, despite the fact that all of these tools accept Ivy or decidable Ivy fragments as inputs. 
We also evaluate our tool and other state of the art tools on a more complex, industrial scale protocol, \textit{MongoLoglessDynamicRaft (MLDR)}~\cite{schultz2021formal}. MLDR performs dynamic reconfiguration in a Raft based replication system. Our tool is the only one which manages to find a correct inductive invariant for MLDR.

To summarize, in this paper we make the following contributions:
\begin{itemize}
    \item A new technique for inductive invariant inference that works for distributed protocols specified in \tlaplus{}. 
    \item A tool, \toolendive{}, which implements our inductive invariant inference algorithm. To our knowledge, this is the only existing tool that works directly on \tlaplus{}.
    \item An experimental evaluation of our tool on a diverse set of distributed protocol benchmarks.
    \item The first, to our knowledge, automatic inference of an inductive invariant for an industrial scale Raft-based reconfiguration protocol. 
\end{itemize}

The rest of this paper is organized as follows. Section~\ref{sec:prelim-and-problem} presents preliminaries and a formal problem statement. Section~\ref{sec:our-approach} describes our algorithm for inductive invariant inference, along with more details on our technique. Section~\ref{sec:evaluation} provides an experimental evaluation of our algorithm, as implemented in our tool, \toolendive{}. Section~\ref{sec:related-work} examines related work, and Section \ref{sec:conclusion-future-work} presents conclusions and goals for future work.

\section{Preliminaries and Problem Statement}
\label{sec:prelim-and-problem}

\newcommand{\varclientlocks}{held}
\newcommand{\varsemaphore}{locked}


\subsubsection{\tlaplus{}} 

Throughout the rest of this paper, we adopt the notation of \tlaplus{}~\cite{lamport2002specifying} for formally specifying systems and their correctness properties. \tlaplus{} is an expressive, high level specification language for specifying distributed and concurrent protocols. It has also been used effectively in industry for specifying and verifying correctness of protocol designs \cite{beersintel08,newcombe2014use}. Note that our tool accepts models written in \tlaplus{}. Figure~\ref{fig:simple-lockserver-example} describes a simple lock server protocol \cite{padon2016ivy,wilcox2015verdi} in \tlaplus{} which we will use as a running example.



\subsubsection{Symbolic Transition Systems}

The protocols considered in this paper can be modeled as parameterized {\em symbolic transition systems} (STSs), like the one shown in Figure~\ref{fig:simple-lockserver-example}.
\remove{This STS is parameterized by two {\em types}, called {\em Server} and {\em Client} (Line 1).
Each type represents a domain of distinct elements, of potentially unbounded, finite size e.g.,
$Server = \{a_1,a_2,a_3,...,a_k\}$ and $Client = \{c_1,c_2,c_3,...,c_k\}$.}
\add{
This STS is parameterized by two {\em sorts}, called {\em Server} and {\em Client} (Line 1).
Each sort represents an uninterpreted constant symbol that can be interpreted as any set of values. In this paper we assume that sorts may only be interpreted over finite domains of distinct values e.g.
$Server = \{a_1,\dots,a_k\}$ and $Client = \{c_1,\dots,c_k\}$.
}

In addition to types, a STS also has a set of {\em state variables}. 
A {\em state} is an assignment of values to all state variables.
We use the notation $s\models P$ to denote that state $s$ {\em satisfies} state predicate $P$, i.e., that $P$ evaluates to true once we replace all state variables in $P$ by their values as given by $s$.

The STS of Figure~\ref{fig:simple-lockserver-example} has two state variables, called {\em \varsemaphore} and {\em \varclientlocks} (Line~2).
The state predicate {\em Init} specifies the possible values of the state variables at an {\em initial state} of the system (Lines~3-5).
{\em Init} states that initially $\varsemaphore[i]$ is {\sc true} for all $i\in Server$, and that $\varclientlocks[i]$ is $\{\}$ (the empty set) for all $i \in Client$. The predicate {\em Next} defines the {\em transition relation} of the STS (Lines~14-16). 
In \tlaplus{}, {\em Next} is typically written as a disjunction of {\em actions} i.e., possible symbolic transitions.
In the example of Figure~\ref{fig:simple-lockserver-example} there are two possible symbolic transitions:
either some client $c$ and some server $s$ engage in a ``connect'' action defined by the {\em Connect}$(c,s)$ predicate,
or some client $c$ and some server $s$ engage in a ``disconnect'' action defined by the {\em Disconnect}$(c,s)$ predicate.


Given two states, $s$ and $s'$, we use the notation $s \to s'$ to denote that there exists a transition from $s$ to $s'$, i.e., that the pair $(s,s')$ satisfies the transition relation predicate $Next$.
A \textit{behavior} is an infinite sequence of states $s_0,s_1,\dots$, such that $s_0 \models Init$ and $s_i\to s_{i+1}$ (i.e., $(s_i,s_{i+1}) \models Next$) for all $i \geq 0$.
A state $s$ is {\em reachable} if there exists a behavior $s_0,s_1,\dots$, such that $s=s_i$ for some $i$. We use $Reach(M)$ to denote the reachable states of a transition system $M$.

The entire set of behaviors of the system is defined as a single temporal logic formula {\em Spec} (Line 17).
In \tlaplus{}, {\em Spec} is typically defined as the \tlaplus{} formula $Init \land {\Box} [ Next ]_{Vars}$, where $\Box$ is the ``always'' operator of linear temporal logic, and $[ Next ]_{Vars}$ represents a transition which either satisfies $Next$ or is a {\em stuttering} step, i.e., where all state variables in $Vars$ remain unchanged.

\subsubsection{Invariants}

In this paper we are interested in the verification of safety properties, and in particular {\em invariants}, which are state predicates that hold at all reachable states. 
Formally, a state predicate $P$ is an \textit{invariant} if $s \models P$ holds for every reachable state $s$. 
The model of Figure~\ref{fig:simple-lockserver-example} contains one such candidate invariant, specified by the predicate {\em Safe} (Line~18). {\em Safe} states that there cannot be two different clients $ci$ and $cj$ which both hold locks to the same server.


\begin{figure}
    \small
    \input{figures/lockserver-tla-example.tex}
    \caption{A simple parameterized protocol defined in \tlaplus{}.}
    \label{fig:simple-lockserver-example}
\end{figure}

\subsubsection{Verification}

The verification problem consists in checking that a system satisfies its specification. In \tlaplus{}, both the system and the specification are written as temporal logic formulas. Therefore, expressed in \tlaplus{}, the safety verification problem we consider in this paper consists of checking that the temporal logic formula 
\begin{align}
    \label{eq_verif_problem}
    Spec \Rightarrow \Box \safe
\end{align}
is {\em valid} (i.e., true under all assignments). That is, establishing that $\safe$ is an invariant of the system defined by $Spec$.


\subsubsection{Finite State Instances}


{\em Instantiating} a \remove{type}\add{sort} means fixing it to a {\em finite} domain of distinct elements. 
For example, we can instantiate {\em Server} to be the set $\{a_1,a_2\}$ (meaning there are only two servers, denoted $a_1$ and $a_2$), and {\em Client} to be the set $\{c_1,c_2\}$ (meaning there are only two clients, denoted $c_1$ and $c_2$).
For the parameterized symbolic transition systems considered in this paper, when we instantiate all \remove{types}\add{sorts} of an STS, the system becomes finite-state, i.e., the set of all possible system states is finite. 

\subsubsection{Inductive Invariants}

A standard technique for solving the safety verification problem~(\ref{eq_verif_problem}) is to come up with an {\em inductive invariant}~\cite{mannasafetybook}. That is, a state predicate $Ind$ which satisfies the following conditions:
\begin{align}
    &Init \Rightarrow Ind \label{cond-init}\\ 
    &Ind \wedge Next \Rightarrow Ind' \label{cond-next}\\
    &Ind \Rightarrow \safe \label{cond-strengthen}
\end{align}
where $Ind'$ denotes the predicate $Ind$ where state variables are replaced by their primed, next-state versions.
Conditions~(\ref{cond-init}) and~(\ref{cond-next}) are, respectively, referred to as \textit{initiation} and \textit{consecution}. 
Condition~(\ref{cond-init}) states that $Ind$ holds at all initial states.
Condition~(\ref{cond-next}) states that $Ind$ is {\em inductive}, i.e., if it holds at some state $s$ then it also holds at any successor of $s$.
Together these two conditions imply that $Ind$ is also an invariant, i.e., that it holds at all reachable states.
Condition~(\ref{cond-strengthen}) states that $Ind$ is stronger than the invariant $\safe$ that we are trying to prove.
Therefore, if all reachable states satisfy $Ind$, they also satisfy $\safe$, which establishes~(\ref{eq_verif_problem}).
The difficulty is in coming up with an inductive invariant which satisfies the above conditions.
The problem we consider in this paper is to infer such an inductive invariant automatically.

\subsubsection{Lemma Invariants}


An inductive invariant $Ind$ typically has the form $Ind \triangleq \safe \wedge A_1 \wedge \dots \wedge A_k$,
where the conjuncts $A_1,...,A_k$ are state predicates and we refer to them as \textit{lemma invariants}. 
Observe that each $A_i$ must itself be an invariant.
The reason is that $Ind$ must be an invariant, i.e., must contain all reachable states, and since $Ind$ is stronger than (i.e., contained in) each $A_i$, each $A_i$ must itself contain all reachable states. Furthermore, although all lemma invariants must be invariants, they need not be individually inductive. However, the conjunction of all lemma invariants together with the safety property $\safe$ must be inductive.
Figure \ref{fig:simple-lockserver-example-indinv} provides an example of an inductive invariant, $Ind$, for the protocol and safety property given in Figure \ref{fig:simple-lockserver-example}. $Ind$ contains a single lemma invariant, $A_1$.
\begin{figure}
    \small
    \input{figures/lockserver-tla-example-indinv.tex}
    \caption{A lemma invariant, $A_1$, and an inductive invariant, $Ind$, for the protocol and safety property given in Figure \ref{fig:simple-lockserver-example}.}
    \label{fig:simple-lockserver-example-indinv}
\end{figure}

\subsubsection{Counterexamples to Induction}

Given a state predicate $P$ (which is typically a candidate inductive invariant), a \textit{counterexample to induction (CTI)} is a state $s$ such that: (1) $s \models P$; and (2) $s$ can reach a state satisfying $\neg P$ in $k$ steps, i.e. there exist transitions $s \to s_1 \to s_2 \to \cdots \to s_k$ and $s_k\models \neg P$. 
That is, a CTI is a state $s$ which proves that $P$ is not inductive i.e., not ``closed'' under the transition relation.
We denote the set of all CTIs of predicate $P$ by $\ctis(P)$. Note that for any inductive invariant $Ind$, the set $\ctis(Ind)$ is empty.
Given another state predicate $Q$ and a state $s \in \ctis(P)$, we say that $Q$ \textit{eliminates} $s$ if $s \not\models Q$, i.e., if $s\models\neg Q$. 


\section{Our Approach}
\label{sec:our-approach}

\begin{figure}[t]
\begin{center}
    \resizebox{.43\textwidth}{!}{
        \begin{tikzpicture}
    \small
    

    \tikzset{osmlogEntryStyle/.style={draw=black,rectangle,minimum width=1.5cm,minimum height=0.5cm,fill=gray!20,thick}}

    \node[draw=black,
    rectangle,
    rounded corners,
    align=center,
    minimum height=5.6cm,
    minimum width=10.6cm,
    line width=1.35,
    fill=gray!5
    ] (outer-box) at (3.5,-0.25){};

    \node[draw=none,
        rectangle, minimum size=1.0cm, 
        text width=1.1cm,
        minimum height=0.5cm,
        fill=white!20, align=center
    ] (input-safety) at (-1.5,3.7){Inputs};

    \node[draw=black,
        rectangle, minimum size=1.0cm, 
        text width=1.1cm,
        minimum height=0.5cm,
        fill=white!20, align=center
    ] (input-spec) at (-1.4,3){$Spec$};

    \node[draw=black,
        rectangle, minimum size=1.0cm, 
        text width=1.1cm,
        minimum height=0.5cm,
        fill=white!20, align=center
    ] (input-safety) at (0.2,3){$Safe$};

    \node[draw,
        rectangle,
        minimum width=1.6cm,
        minimum height=1.0cm,
        text width=2cm,
        fill=blue!10,
        align=center
    ] (inv-gen) at (0.5,0){Invariant Generator};


    \node[draw,
        cylinder,
        rotate=90,
        minimum size=1.0cm,
        minimum height=1.2cm,
        fill=gray!30
    ] (inv-db) at (2.0,1.5){};

    \node[draw=none,
        circle,
        rotate=0,
        minimum size=1.0cm,
        minimum height=1.2cm,
        fill=none
    ] (inv-db-text) at (inv-db){$Invs$};

    \node[draw=black,rectangle,
        minimum width=2.7cm, minimum height=3.85cm, text width=2cm,
        fill=gray!20,align=center
    ] (ctibox) at (3.7,-1){};

    \node[draw,
    rectangle,
    align=center,
    minimum size=1.5cm,
    text width=1.55cm,
    minimum height=1.2cm,
    minimum width=2.1cm,
    fill=orange!20
    ] (cti-elim) at (3.7,0){CTI Eliminator};

    \node[draw,
    rectangle,
    align=center,
    minimum size=1.5cm,
    text width=1.55cm,
    minimum height=1.2cm,
    minimum width=2.1cm,
    fill=cyan!20
    ] (cti-elim) at (3.7,0){CTI Eliminator};

    \node[draw,
    rectangle,
    align=center,
    minimum size=1.0cm,
    text width=1.55cm,
    minimum height=1.2cm,
    minimum width=2.2cm,
    fill=cyan!15
    ] (cti-gen) at (3.7,-2.1){CTI Generator};

    \node[draw=black,rectangle,align=center,
    minimum size=0cm, fill=white!20
    ] (indout) at (7.0,0){$\begin{aligned}
        Ind &\triangleq \\
            \wedge& Safe \\
            \wedge& A_1\\
            \phantom{\wedge}& \vdots\\
            \wedge& A_{k}\\
            \color{blue}{\wedge}& \color{blue}{A_{k+1}}\\
    \end{aligned}$};

    \node[draw=none,circle] (output-node) at (10,0.0){Output};

    \draw[-stealth] (inv-gen) to [bend left=20] node[above left]{$I_{new}$} (inv-db);

    \draw[-stealth] (inv-db) to [bend left=20] node[above right]{$I \subseteq Invs$} (cti-elim);

    \draw[-stealth] (cti-gen) to [bend right=0] node[right]{$CTIs$} (cti-elim);

    
    \draw[-stealth] (cti-elim) -- (indout) node[midway,above]{$\color{blue}{A_{k+1}}$};

    \draw[-stealth] (indout) |- (cti-gen);
    
    \draw[-stealth] (input-spec) |- (inv-gen) node[midway,above]{};
    \draw[-stealth] (input-spec) |- (cti-gen) node[midway,above]{};
    \draw[-stealth] (input-safety) -| (indout) node[midway,above]{};
    \draw[-stealth] (indout) -- (output-node) node[midway,above]{};

\end{tikzpicture}     
    }       
\end{center}
\caption{Components of our technique for inductive invariant inference.}
\label{fig:block-diagram}
\end{figure}
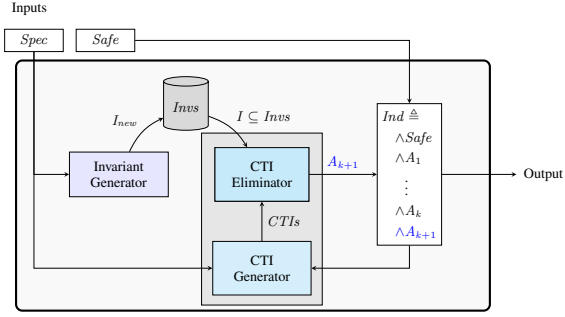


At a high level, our inductive invariant inference method consists of the following steps:
\begin{enumerate}
\item Generate many candidate lemma invariants, and store them in a repository that we call $\invs$.
\item Generate counterexamples to induction for a current candidate inductive invariant, $Ind$.
 If we cannot find any such CTIs, return $Ind$.
\item Select lemma invariants from $\invs$ so that all CTIs are eliminated. If we cannot eliminate all CTIs, either give up,
or go to Step~1 and populate the repository with more lemma invariants. Otherwise, add the selected lemma invariants to $Ind$ and repeat from Step~2.
\end{enumerate}

The conceptual approach is illustrated in Figure~\ref{fig:block-diagram}.
Our detailed algorithm is described in Section~\ref{sec:inv-inference-algo}.
Section~\ref{sec:invariant-generation} provides details on our lemma invariant generation procedure, Section~\ref{sec:CTIgeneration} provides details on CTI generation, and Section~\ref{sec:lemma-inv-selection} describes the selection of lemma invariants.

\subsection{Inductive Invariant Inference Algorithm}
\label{sec:inv-inference-algo}

Our inductive invariant inference algorithm is given in pseudocode in Algorithm \ref{fig:invgen-algo}.
The algorithm takes as input:
(1) a finite instance of a symbolic transition system $M$,
(2) a candidate invariant (safety property) $\safe$, (3) a lemma invariant repository $\invs$, and (4) a grammar $G$ for generating lemma invariant candidates. The use of the grammar is discussed further in Section \ref{sec:invariant-generation}. $\invs$ may initially be empty, or be pre-populated from previous runs of the algorithm.
The algorithm aims to discover an inductive invariant, $Ind$, of the form $Ind = Safe \wedge A_1 \wedge \dots \wedge A_n$. 


\begin{algorithm}[t]
    \caption{Our inductive invariant inference algorithm.}
    \label{fig:invgen-algo}
    \footnotesize
\algnewcommand{\algorithmicgoto}{\textbf{goto}}%
\algnewcommand{\Goto}[1]{\algorithmicgoto~Line \ref{#1}}%
\begin{algorithmic}[1]
    \State \textbf{Inputs:} 
    \Statex {$M$: Finite instance of a parameterized STS} 
    \Statex {$\safe$: Candidate invariant} 
    \Statex {$\invs$: Lemma invariant repository (typically empty initially)} 
    \Statex {$G$: Grammar for invariant generation} 
    \vspace{0.6mm}
    \Procedure{InferInductiveInvariant}{$M$, $\safe$, $G$, $\invs$}
    \State $Ind \gets \safe$   \label{algolinesafety}
    \State $X \gets GenerateCTIs(M, Ind)$   \label{algolineCTIsinit}
    \State $\invs \gets GenerateLemmaInvariants(M, \invs, G)$  \label{line:gen-invs}
    %
    \While{$X \neq \emptyset$} \label{algolineloopbegin}
    \If{$\exists A \in \invs :A$ eliminates at least one CTI in $X$}
		\State pick $A_{max} \in \invs$ that eliminates the most CTIs from $X$ 
		\State $Ind \gets Ind \wedge A_{max}$    \label{algolineIndupdate}
		\State $X \gets X \setminus \{s \in X : s \not\models A_{max}\}$  \label{algolineXupdate}
    \Else \label{algolinecannoteliminate}
		\State either \Goto{line:gen-invs}    \label{algolinegoto}
        \State or \Return ($Ind$, ``Fail: couldn't eliminate all CTIs.'')  \label{invgen-algo:line-fail}
    \EndIf
    \State $X \gets GenerateCTIs(M, Ind)$    \label{algolineCTIsnew}
    \EndWhile
    \State \Return ($Ind$, ``Success: managed to eliminate all CTIs.'')  \label{algolinesuccess}
    \EndProcedure
\end{algorithmic}

\end{algorithm}

The algorithm maintains a current inductive invariant candidate, $Ind$, which it initializes to $\safe$, the safety property that we are trying to prove (Line~\ref{algolinesafety}).
It then generates a set $X$ of CTIs of $Ind$ (Line~\ref{algolineCTIsinit}).
The algorithm may also initialize the repository of lemma invariants, $\invs$, or add more lemma invariants to $\invs$ if it is initially non-empty (Line~\ref{line:gen-invs}).
The procedures $GenerateLemmaInvariants$ and $GenerateCTIs$ are described in more detail below, in
Sections~\ref{sec:invariant-generation} and~\ref{sec:CTIgeneration}, respectively.

In its main loop, the algorithm tries to eliminate all currently known CTIs. 
As long as the set $X$ of currently known CTIs is non-empty, the algorithm tries to find a lemma invariant in the $\invs$ repository that eliminates the maximal number of remaining CTIs possible. If such a lemma invariant exists, the algorithm adds it as a new conjunct to $Ind$ (Line~\ref{algolineIndupdate}), removes from $X$ the CTIs that were eliminated by the new conjunct (Line~\ref{algolineXupdate}), and proceeds by attempting to generate more CTIs, since the updated $Ind$ is not necessarily inductive (Line~\ref{algolineCTIsnew}).

If no lemma invariant exists in the current repository $\invs$ that can eliminate any of the currently known CTIs (Line~\ref{algolinecannoteliminate}), then we may either (1) generate more lemma invariants in the repository, or (2) give up. The first choice is implemented by the {\bf goto} statement in Line~\ref{algolinegoto}.
The second choice represents a failure of the algorithm to find an inductive invariant (Line~\ref{invgen-algo:line-fail}).
However, in this case we still return $Ind$ since, even though it is not inductive, it may contain several useful lemma invariants. These lemma invariants are useful in the sense that they might be part of an ultimate inductive invariant.

If all known CTIs have been eliminated, the algorithm terminates successfully and returns $Ind$ (Line~\ref{algolinesuccess}).
Successful termination of the algorithm indicates that the returned $Ind$ is likely to be inductive.
However
our method does not provide a formal inductiveness guarantee. $Ind$ might not be inductive for a number of reasons. First, as we discuss further in Section~\ref{sec:CTIgeneration}, our CTI generation procedure is probabilistic in nature, and therefore $GenerateCTIs$ might miss some CTIs. Second, even if the finite-state instance $M$ explored by the algorithm has no remaining CTIs, there might still exist CTIs in other instances of the \sts, for larger parameter values.

Even though a candidate invariant returned by a successful termination of Algorithm~\ref{fig:invgen-algo} is not formally guaranteed to be inductive, we ensure soundness of our overall procedure by doing a final check that the discovered candidate inductive invariant is correct using the \tlaplus{} proof system (TLAPS)~\cite{cousineau2012tla}. Validation of invariants in TLAPS is discussed further in Section~\ref{sec:ind-ind-validation}.
In practice we found that all of the invariants generated in our evaluation (Section \ref{sec:evaluation}) are correct inductive invariants. 


We also remark that in the current version of our algorithm and in the current implementation of our tool, we only explore the single finite-state instance of the \sts provided by the user, and we do not attempt to automatically increase the bounds of the parameters within the algorithm, as is done for example in the approach described in \cite{goel2021symmetry}. This is, however, a relatively straightforward extension to our algorithm, and would like to explore this option in future work.

\subsection{Lemma Invariant Generation}
\label{sec:invariant-generation}

For a given finite instance $M$ of a parameterized transition system, the goal of lemma invariant generation is to produce a set of state predicates that are invariants of $M$. To search for these invariants, we adopt an approach similar to other, \textit{syntax-guided synthesis} based techniques \cite{Fedyukovichsyguinvs,2018accelsynsyninvs} for invariant discovery. We randomly sample invariant candidates from a defined \textit{grammar}, which is  generated from a given set of $seed$ predicates. Each seed predicate is an atomic boolean predicate over the state variables of the system. Note that the parameterized distributed protocols that we consider in this paper typically have inductive invariants that are universally or existentially quantified over the parameters of the protocol or other values of the system state. So, our invariant generation technique assumes a fixed quantifier template that is provided as input. The provided seed predicates are unquantified predicates that can contain bound variables that appear in the given quantifier template. An example of a simple grammar for the protocol of Figure \ref{fig:simple-lockserver-example} is shown in Figure \ref{fig:lockserver-grammar-seeds}. 

\begin{figure}
    \small
    \input{figures/lockserver-grammar-seeds-example.tex}
    \caption{Example of a grammar for lemma invariant generation for the \textit{lockserver} protocol shown in Figure \ref{fig:simple-lockserver-example}. The list of unquantified \textit{seed} predicates and the quantifier template, \textit{quant}, are provided as user inputs.}
    \label{fig:lockserver-grammar-seeds}
\end{figure}

Candidate invariants are produced by generating random predicates over the space of seed predicates. Specifically, a candidate predicate is formed as a random disjunction of seed predicates, where each disjunct may be negated with probability $\frac{1}{2}$. The logical connectives $\{\vee,\neg\}$ are functionally complete \cite{Wernick1942CompleteSO}, so they serve as a simple basis for generating candidate invariants, which we chose to reduce the invariant search space.

For a given set of candidate invariants, $C$, we check which of the predicates in $C$ are invariants using an explicit state model checker. This can be done effectively due to our use of the small scope hypothesis i.e. the fact that we reason only about a finite instance $M$ of a  parameterized transition system. This largely reduces the invariant checking problem to a  data processing task. Namely: 
\begin{enumerate}[(1)]
    \item Generate $Reach(M)$, the set of reachable states of $M$.
    \item Check that $s \models P$ for each predicate $P \in C$ and each $s \in Reach(M)$.
\end{enumerate}
Note that after (1) has been completed once, the set of reachable states can be cached and only step (2) must be re-executed when searching for additional invariants.

In theory, the worst case cost of step (2) is proportional to $|C| \cdot |Reach(M)|$. In practice, however, it can often be much less costly than this, since once a state violates a predicate $P$, $P$ need not be checked further. Furthermore, both of the above computation steps are highly parallelizable, a fact we make use of in our implementation, as discussed further in Section~\ref{sec:impl-and-experiment-setup}.

We also remark that, in practice, the $GenerateLemmaInvariants$ procedure is configured to search for candidate invariants of a fixed term size i.e. with a fixed or maximal number of disjuncts. In our implementation, presented in Section~\ref{sec:impl-and-experiment-setup}, we utilize this to search for smaller invariants (fewer terms) first, before searching for larger ones. That is, we prefer to eliminate CTIs if possible with smaller invariants before searching for larger ones. This aims to bias our procedure towards discovery of compact inductive invariant lemmas. 

Furthermore, since $GenerateLemmaInvariants$ does not employ an exhaustive
search for invariants over a given space of predicates, it accepts a numeric parameter, $\nlemmasparam$, which determines how many candidate predicates to sample. More details of how the concrete values of this parameter are configured are discussed in our evaluation, in Section \ref{sec:evaluation}.

\subsection{CTI Generation}
\label{sec:CTIgeneration}

Each round of our algorithm relies on access to a set of multiple CTIs, as a means to prioritize between different choices of new lemma invariants. To generate these CTIs, we use a probabilistic technique proposed in \cite{Lamport2018UsingTT} that utilizes the TLC explicit state model checker \cite{tlcmodelchecker}. 
Given a finite instance of a STS $M$ with system states $S$, transition relation predicate $Next$, and given candidate inductive invariant $Ind$, the procedure $GenerateCTIs(M, Ind)$ works by calling 
the TLC model checker. TLC attempts to randomly sample states $s_0 \in S$ for which there exists a sequence of states $s_1,s_2,\dots,s_{k-1},s_k \in S$, such that both of the following hold:
\begin{itemize}
    \item $\forall i = 0,1,\dots,k-1 : (s_i,s_{i+1}) \models Next \land s_i \models Ind$
    \item $s_k \not\models Ind$.
\end{itemize}
The model checker will report this behavior, 
and all states $s_0,s_1,s_2,\dots,s_{k-1}$ are recorded as counterexamples to induction.




Due to the randomized nature of this technique, the CTI generation procedure requires a given parameter, $N_{ctis}$, that effectively determines how many possible states TLC will attempt to sample before terminating the CTI generation procedure. This is required, since, for systems with sufficiently large state spaces, even if finite, sampling all possible states is infeasible. Generally, this parameter can be tuned based on the amount of compute power available to the tool, or a latency tolerance of the user. We discuss more details of this parameter and how it is tuned in our experiments in Section \ref{sec:evaluation}. 

In practice, during our evaluation we found that TLC was able to effectively generate many thousands of CTIs at each round of the inference algorithm using the above technique. This provided an adequately diverse distribution of CTIs for effectively guiding our counterexample elimination procedure, which we describe in more detail in Section \ref{sec:lemma-inv-selection}. Section \ref{sec:evaluation} presents more detailed metrics on CTI generation as measured when testing our implementation on a variety of protocol benchmarks. In future we feel it would be valuable to explore and compare with other, SMT/SAT based techniques for this type of counterexample generation task \cite{2018efficientSATSampling,2019tlamadesymbolic}. 



\subsection{Lemma Invariant Selection by CTI Elimination}
\label{sec:lemma-inv-selection}

The task of selecting lemma invariants for use as inductive invariant conjuncts is based on a process of CTI elimination, as described briefly in Section \ref{sec:inv-inference-algo}. That is, CTIs are used as guidance for which invariants to choose for new lemma invariants to append to the current inductive invariant candidate. 
Once a sufficiently large set of CTIs has been generated, as discussed in Section~\ref{sec:CTIgeneration}, we select lemma invariants using a greedy heuristic of CTI elimination, which we describe below.

\subsubsection{CTI Elimination}
\label{sec:cti-elim}

Recall that a CTI $s$ is {\em eliminated} by a state predicate $A$ if $s\not\models A$.
When examining a current set of CTIs, $X$, our algorithm looks for the next lemma invariant $A \in \invs$ that eliminates the most CTIs in $X$. The algorithm
will continue choosing additional lemma invariants according to this strategy until all counterexamples are eliminated, or until it cannot eliminate any further counterexamples. Each selected invariant $A_i \in Invs$ will be appended as a new conjunct to the current inductive invariant candidate i.e. $Ind \gets Ind \wedge A_i$. Once all counterexamples have been eliminated, the tool will terminate and return a final candidate inductive invariant. This is a simple heuristic for choosing new invariant conjuncts that aims to bias the overall inductive invariant towards being relatively concise. That is, if we have a choice between two alternate lemma conjuncts to choose from, we prefer the conjunct that eliminates more CTIs. 


More generally, lemma selection at each round of the algorithm can be viewed as a version of the set covering problem \cite{clrsthirded}. 
Ideally,
we would like to find the smallest set of lemma invariants that eliminate (i.e. cover) the set of CTIs $X$. Solving this problem optimally is known to be NP-complete \cite{Karp1972}, but we have found a greedy heuristic \cite{1979Chvatalgreedysetcover} to work sufficiently well in our experiments, the results of which are presented in Section \ref{sec:evaluation}. 
In future we would like to explore more sophisticated heuristics for lemma selection that take into account additional metrics, like syntactic invariant size, quantifier depth, etc.  

\subsection{Validation of Inductive Invariant Candidates}
\label{sec:ind-ind-validation}


If our inference algorithm terminates successfully, it will return a candidate inductive invariant. Since we look for invariants on finite protocol instances, though, this candidate may not be an inductive invariant for general (e.g. unbounded) protocol instances.
So, upon termination, we check to see if the returned candidate invariant is truly inductive for all protocol instances by passing it to an SMT solver. Currently, we use the \tlaplus{} proof system (TLAPS) \cite{cousineau2012tla} for this step, which generates an SMT encoding for \tlaplus{} \cite{2016merzmanysorted}.

For many of the protocols we tested and the invariants discovered by our tool, we found that this step was fully automated (see Section~\ref{sec:evaluation} and Table~\ref{fig:benchmark-results-detailed-extra} in the Appendix). That is, no user assistance was required to establish validity of the discovered invariant. In cases where the underlying solver cannot automatically prove the candidate inductive invariant, some amount of human guidance can be provided by decomposing the proof into smaller SMT queries. We have completed this validation step for all of the inductive invariant candidates discovered in our experiments, and we confirmed that all candidate invariants produced by our tool were indeed correct inductive invariants (see Section~\ref{sec:evaluation}).

\section{Implementation and Evaluation}
\label{sec:evaluation}




 \subsection{Implementation and Experimental Setup}
 \label{sec:impl-and-experiment-setup}

 Our invariant inference algorithm is implemented in a tool, \toolendive{}, whose main implementation consists of approximately 2200 lines of Python code. There are also some optimized subroutines which consist of an additional few hundred lines of C++ code.
 Internally, \toolendive{} makes use of version 2.15 of the TLC model checker \cite{tlcmodelchecker}, with some minor modifications to improve the efficiency of checking many invariants simultaneously. TLC is used by \toolendive{} for most of the algorithm's compute intensive verification tasks, like checking candidate lemma invariants (Section~\ref{sec:invariant-generation}) and CTI elimination checking (Section~\ref{sec:cti-elim}). 
 
 For all of the experiments discussed below, \toolendive{} is configured to use 24 parallel TLC worker threads for invariant checking, 4 parallel threads for CTI generation, and 4 threads for CTI elimination. CTI generation and CTI elimination can be parallelized further in a straightforward manner, but we limit these procedures to 4 parallel threads to simplify certain aspects of our current implementation. 

 For each benchmark run, we initialize $Invs$ (as explained in Algorithm \ref{fig:invgen-algo}) as an empty set and configure the lemma invariant generation procedure discussed in Section~\ref{sec:invariant-generation} with a parameter value of $\nlemmasparam=15000$. The grammars used for invariant generation were mined from predicates appearing in each protocol specification.
 
 We configure our CTI generation procedure with a parameter value of $N_{ctis}=50000$. 
 $N_{ctis}$ does not directly correspond to how many concrete CTI states will be generated, but a higher value indicates TLC will sample more states when searching for CTIs. 
 We also limit the maximum number of CTIs returned by each call to the $GenerateCTIs$ procedure to 10000 states. In theory, generating more CTIs provides better counterexample diversity, and is therefore  better for our CTI elimination heuristics. We impose an upper limit, however, to avoid scalability issues in our tool's current implementation. In practice we found this limit sufficient to provide effective guidance for lemma invariant selection.
 
 All of our experiments were run on a 48-core Intel(R) Xeon(R) Gold 5118 CPU @ 2.30GHz machine with 196GB of RAM. 


 \subsection{Benchmarks}
 \label{sec:benchmarks}

To evaluate \toolendive{}, we measured its performance on \numbenchmarks{} protocols selected from an existing benchmark set published in \cite{goel2021symmetry}. We also evaluate endive on an additional, industrial scale protocol, \textit{MongoLoglessDynamicRaft (MLDR)}, which is a recent protocol for distributed dynamic reconfiguration in a Raft based replication system \cite{schultz2021design,schultz2021formal}.


\subsubsection{Protocol Conversion}

The \numbenchmarks{} benchmarks we used from \cite{goel2021symmetry} were originally specified in Ivy \cite{padon2016ivy}, but \toolendive{} accepts protocols in \tlaplus{}, so it was necessary to manually translate the protocols from Ivy to \tlaplus{}. There are significant differences in how protocols are specified in Ivy and \tlaplus{}. The underlying approach to modeling systems as discrete transition systems, however, by specifying initial states and a transition relation, are common between them. In our manual translation, we aimed to emulate the original Ivy model as close as possible.

The formal specification for the \textit{MongoLoglessDynamicRaft} protocol (\textit{MLDR}) was originally written in \tlaplus{} \cite{schultz2021design}.
Thus, in order to compare with other invariant inference tools which accept Ivy as their input language, we had to translate MLDR from \tlaplus{} into Ivy.
This conversion process was highly nontrivial due to the significant differences between the Ivy and \tlaplus{} languages. \tlaplus{} is a very expressive language that includes integers, strings, sets, functions, records, and sequences as primitive data types along with their standard semantics. In contrast, the Ivy modeling language, RML \cite{padon2016ivy}, includes only basic, first order relations and functions. For more complex datatypes (e.g. arrays or sequences), their semantics must be defined and axiomatized manually.

An artifact containing all of our source code and instructions for reproducing our evaluation results can be found at \cite{fmcad22artifact}. A public, open-source version of our tool is also available at \cite{endivetool}.





\subsection{Results}
\label{sec:results-discussion}
Our overall results are shown in Table~\ref{fig:benchmark-results}. We compared \toolendive{} with four recent, state of the art techniques for inferring invariants of distributed protocols: IC3PO \cite{goel2021symmetry}, fol-ic3 \cite{2020firstorderquantified}, SWISS \cite{hance2021finding}, and DistAI \cite{yao2021distai}. Note that \toolendive{} accepts protocols in \tlaplus{}, whereas all other tools accept protocols in Ivy or mypyvy.


The numbers shown for both IC3PO and fol-ic3 in Table~\ref{fig:benchmark-results} are as reported in the evaluation presented in \cite{goel2021symmetry}, with timeouts indicated by a \textit{TO} entry. For the SWISS results in Table~\ref{fig:benchmark-results}, where possible, we show the runtime numbers reported in \cite{hance2021finding}, indicated with a $\dagger$ mark. For the benchmarks in Table~\ref{fig:benchmark-results} that were not tested in \cite{hance2021finding}, we present the results from our own runs of the tool, all using default SWISS configuration parameters. We ran SWISS both with an invariant template matching our own template for \toolendive{} and also in automatic mode, and report the better of the two results. The results for DistAI are reported from our runs using the tool in its default configuration. For DistAI and SWISS, we report an \textit{err} result in cases where the tool returned an error without producing a result.
We report a \textit{fail} result in cases where DistAI or SWISS terminated without error but did not discover an inductive invariant. In all cases where a benchmark protocol was not available in the required input language for the corresponding tool, we mark this with an \textit{n/a} entry.

For each benchmark result in Table \ref{fig:benchmark-results}, we report the total wall clock time to discover an inductive invariant in the \textit{Time} column, along with the number of total lemma invariants contained in the discovered invariant, including the safety property, in the \textit{Inv} column. 
Note that the number of total lemmas in the invariants discovered by SWISS was not reported in \cite{hance2021finding}. Thus, we report the number of lemmas discovered by SWISS in our own runs, for the cases where we were able to run SWISS successfully to produce an invariant.

More detailed statistics on the \toolendive{} benchmark results are provided in Appendix \ref{sec:appendix-results}, specifically: the number of eliminated CTIs, runtime profiling information, finite instance sizes used, and automation level of the TLAPS proofs.

\begin{table}[t]
    \begin{center}
        \begingroup
        \setlength{\tabcolsep}{1.2pt} 
        \renewcommand{\arraystretch}{1.0} 
        \scriptsize
        \begin{tabular}{ l | l | r r | r r | r r | r r | r r  }
\multicolumn{2}{c}{} & \multicolumn{2}{c}{\textbf{endive}} & \multicolumn{2}{c}{IC3PO} & \multicolumn{2}{c}{fol-ic3} & \multicolumn{2}{c}{SWISS} & \multicolumn{2}{c}{DistAI}\\\hline
No. & Protocol & Time & Inv & Time & Inv & Time & Inv & Time & Inv & Time & Inv\\ \hline 
1 & tla-consensus & 1 & 1 & 0 & 1 & 1 & 1 & 1 & 2 & 2 & 1 \\ 
2 & tla-tcommit & 2 & 1 & 1 & 2 & 2 & 3 & 2 & 8 & 2 & 7 \\ 
3 & i4-lock-server & 7 & 2 & 1 & 2 & 1 & 2 & \textsuperscript{$\dagger$}1 & 2 & \gray{err} &   \\ 
4 & ex-quorum-leader-election & 11 & 2 & 3 & 5 & 24 & 8 & 11 & 5 & 3 & 8 \\ 
5 & pyv-toy-consensus-forall & 19 & 3 & 3 & 5 & 11 & 5 & \textsuperscript{$\dagger$}3 & 7 & \gray{err} &   \\ 
6 & tla-simple & 8 & 2 & 6 & 3 & \gray{TO} &   & 28 & 8 & \gray{err} &   \\ 
7 & ex-lockserv-automaton & 23 & 9 & 7 & 12 & 10 & 12 & \gray{fail} &   & 2 & 13 \\ 
8 & tla-simpleregular & 10 & 4 & 8 & 4 & 57 & 9 & 65 & 21 & \gray{err} &   \\ 
9 & pyv-sharded-kv & 312 & 6 & 10 & 8 & 22 & 10 & \textsuperscript{$\dagger$}4024 &   & 2 & 16 \\ 
10 & pyv-lockserv & 35 & 9 & 11 & 12 & 8 & 11 & \textsuperscript{$\dagger$}3684 &   & 2 & 13 \\ 
11 & tla-twophase & 43 & 10 & 14 & 9 & 9 & 12 & 33 & 24 & 29 & 306 \\ 
12 & i4-learning-switch & \gray{TO} &  & 14 & 10 & \gray{TO} &   & \gray{TO} &   & 21 & 32 \\ 
13 & ex-simple-decentralized-lock & 44 & 4 & 19 & 15 & 4 & 8 & 1 & 2 & 26 & 17 \\ 
14 & i4-two-phase-commit & 69 & 11 & 27 & 11 & 8 & 9 & \textsuperscript{$\dagger$}6 & 15 & 17 & 67 \\ 
15 & pyv-consensus-wo-decide & 127 & 8 & 50 & 9 & 168 & 26 & \textsuperscript{$\dagger$}18 & 8 & \gray{err} &   \\ 
16 & pyv-consensus-forall & 175 & 8 & 99 & 10 & 2461 & 27 & \textsuperscript{$\dagger$}29 & 9 & \gray{err} &   \\ 
17 & pyv-learning-switch & \gray{TO} &  & 127 & 13 & \gray{TO} &   & \textsuperscript{$\dagger$}959 &   & 79 & 70 \\ 
18 & i4-chord-ring-maintenance & \gray{n/a} &  & 229 & 12 & \gray{TO} &   & \textsuperscript{$\dagger$}\gray{TO} &   & 53 & 164 \\ 
19 & pyv-sharded-kv-no-lost-keys & 13 & 2 & 3 & 2 & 3 & 2 & \textsuperscript{$\dagger$}1 & 4 & \gray{fail} &   \\ 
20 & ex-naive-consensus & 40 & 4 & 6 & 4 & 73 & 18 & 18 & 5 & \gray{fail} &   \\ 
21 & pyv-client-server-ae & 46 & 2 & 2 & 2 & 877 & 15 & \textsuperscript{$\dagger$}3 & 5 & \gray{err} &   \\ 
22 & ex-simple-election & 24 & 4 & 7 & 4 & 32 & 10 & 9 & 5 & \gray{err} &   \\ 
23 & pyv-toy-consensus-epr & 19 & 4 & 9 & 4 & 70 & 14 & \textsuperscript{$\dagger$}2 & 4 & \gray{err} &   \\ 
24 & ex-toy-consensus & 7 & 2 & 10 & 3 & 21 & 8 & 6 & 4 & \gray{err} &   \\ 
25 & pyv-client-server-db-ae & 4941 & 8 & 17 & 6 & \gray{TO} &   & \textsuperscript{$\dagger$}24 & 13 & \gray{err} &   \\ 
26 & pyv-hybrid-reliable-broadcast & \gray{n/a} &  & 587 & 4 & 1360 & 23 & \textsuperscript{$\dagger$}\gray{TO} &   & \gray{err} &   \\ 
27 & pyv-firewall & 38 & 5 & 2 & 3 & 7 & 8 & 75 & 5 & \gray{err} &   \\ 
28 & ex-majorityset-leader-election & 53 & 4 & 72 & 7 & \gray{TO} &   & 28 & 10 & \gray{err} &   \\ 
29 & pyv-consensus-epr & 247 & 8 & 1300 & 9 & 1468 & 30 & 72 & 10 & \gray{err} &   \\ 
\hline
30 & mldr & 2025 & 6 & \gray{TO} &   & \gray{n/a} &   & \gray{err} &   & \gray{err} &   \\ 
\end{tabular}

        \endgroup
    \end{center}
    \caption{Distributed protocol benchmark results.}
    \label{fig:benchmark-results}
\end{table}

\subsection{Comparison with Other Tools}

Although Table~\ref{fig:benchmark-results} relates our approach to several others, we note that our tool is not directly comparable to other tools. The most fundamental difference is that our tool accepts \tlaplus{} whereas all other tools in Table~\ref{fig:benchmark-results} accept Ivy or mypyvy. Furthermore, some tools work only with the restricted decidable EPR or extended EPR fragments of Ivy. To our knowledge, this is the case with SWISS and DistAI. As a result, our tool is a-priori less automated than other tools, following a standard tradeoff between expressivity and automation. In practice, however, and despite this theoretical limitation, our tool produces a result in most cases, while some of the a-priori more automated tools time out or fail.

Another important difference between the tools of Table~\ref{fig:benchmark-results} is what kind of inductive invariants can be produced by each tool. In our case, the user provides the grammar of possible lemma invariants as an input to the tool, allowing both universal and existentially quantified invariants ($\forall$ and $\exists$). DistAI is limited to only universally quantified ($\forall$)
invariants, and SWISS is limited to invariants that fall into the extended EPR fragment, though it can learn both universal and existentially quantified invariants. Both fol-ic3 and IC3PO attempt to learn the quantifier structure itself during counterexample generalization, and can infer both universal and existentially quantified invariants. These tools do not always guarantee, however, that the discovered invariants will fall into a decidable logic fragment. Thus, they provide no explicit guarantee that the overall inference procedure will, in general, be fully automated. 

\subsection{Discussion}

Our tool, \toolendive{}, was able to successfully discover an inductive invariant for \numbenchmarkssolved{} of the \numbenchmarks{} protocol benchmarks from \cite{goel2021symmetry}, and all of the invariants it discovered were proven correct using TLAPS.  For the two protocols out of these 29 that our tool did not solve, pyv-learning-switch and i4-learning-switch, this was due to scalability limitations of CTI generation, which we believe could be improved with a smarter CTI generation algorithm or by incorporating a symbolic model checker \cite{2019tlamadesymbolic} for this task.

\toolendive{} was also able to automatically discover an inductive invariant for a key safety property of \textit{MLDR}, a Raft-based distributed dynamic reconfiguration protocol \cite{schultz2021design}. This protocol, reported in Table \ref{fig:benchmark-results} as \textit{mldr}, is a significantly more complex, industrial scale protocol \cite{schultz2021formal}. 
IC3PO was not able to discover an invariant for our Ivy model of the MLDR protocol after a 1 hour timeout when given the same instance size used in the \tlaplus{} model given to \toolendive{}. SWISS and DistAI both produced an error when run on our Ivy model of MLDR.




Generally, the wall clock time taken for \toolendive{} to discover an inductive invariant is of a similar order of magnitude to IC3PO. \toolendive{} even outperforms IC3PO in some cases, despite the fact that \toolendive{} works with \tlaplus{} and IC3PO works with Ivy. 
%
%
%
Moreover, in several cases where \toolendive{}'s runtime exceeds that of IC3PO, \toolendive{} is able to discover a smaller inductive invariant (e.g. pyv-lockserv, ex-simple-decentralized-lock, pyv-consensus-forall). Additionally, \toolendive{} is often able to discover a considerably smaller invariant than tools like DistAI and SWISS. For example, on tla-twophase, \toolendive{} learns an invariant with 10 overall conjuncts, whereas SWISS learns a 24 conjunct invariant, and DistAI learns a much larger invariant, with over 300 conjuncts. \toolendive{} performs similarly well for the tla-simpleregular and i4-two-phase-commit benchmarks. This demonstrates that \toolendive{} compares favorably against other enumerative approaches for inductive invariant inference, both in terms of efficiency and compactness of invariants, while also working over \tlaplus{}, a much more expressive input language.

It is additionally worth noting that our current \toolendive{} implementation is not highly optimized. In particular, the TLC model checker, used internally by endive, is implemented in Java and interprets \tlaplus{} specifications dynamically \cite{kuppe2017verified}, rather than compiling models to a low level, native representation as done by tools like SPIN \cite{holzmann1997model}. As a result, TLC may not be the most efficient for our inference procedure, and could likely be optimized further.



\section{Related Work}
\label{sec:related-work}

There are several recently published techniques that attempt to solve the problem of inductive invariant inference for distributed protocols.
The IC3PO tool \cite{goel2021symmetry}, which extended the earlier I4 tool \cite{Ma2019}, uses a technique based on IC3 \cite{bradley2011sat} with a novel \textit{symmetry boosting} technique that serves to accelerate IC3/PDR and also to infer the quantifier structure of lemma invariants. 
The fol-ic3 algorithm presented in \cite{2020firstorderquantified} presents another IC3 based algorithm which uses a novel \textit{separators} technique for discovering quantified formulas to separate positive and negative examples during invariant inference. SWISS \cite{hance2021finding} is another recent approach that uses an enumerative search for quantified invariants while using the Ivy tool to validate possible inductive candidates. It relies on SMT based reasoning over an unbounded domain, and does not reason directly about finite instances of distributed protocols. DistAI \cite{yao2021distai} uses a similar approach but additionally utilizes a technique of sampling reachable protocol states to filter invariants, which is similar to our approach of executing explicit state model checking as a means to quickly discover invariants. DistAI is limited, however, to learning only universally quantified invariants.

In addition to these inductive invariant inference techniques, there also exists prior work on alternative techniques for parameterized protocol verification. These include approaches based on cutoff detection \cite{2016abdulla}, regular model checking \cite{2004bouajani}, and symbolic backward reachability analysis \cite{2010mcmt}.

More broadly, there exist many prior techniques for the automatic generation of program and protocol invariants that rely on data driven or grammar based approaches. Houdini \cite{houdiniflanagan} and Daikon \cite{ernst2007daikon} both use enumerative checking approaches to discover program invariants. FreqHorn \cite{2018accelsynsyninvs} tries to discover quantified program invariants about arrays using an enumerative approach that discovers invariants in stages and also makes use of the program syntax. Other techniques have also tried to make invariant discovery more efficient by using improved search strategies based on MCMC sampling \cite{sharma2016invariant}.

\section{Conclusions and Future Work}
\label{sec:conclusion-future-work}

We presented a new technique for inferring inductive invariants for distributed protocols specified in \tlaplus{}
and evaluated it on a diverse set of protocol benchmarks. 
Our approach is novel in that: (1) it is the first, to our knowledge, to infer inductive invariants directly for protocols specified in \tlaplus{} and (2) it is based around a core procedure for generating plain, not necessarily inductive, lemma invariants.
%
Our results show that our approach performs strongly on a diverse set of distributed protocol benchmarks. In addition, it is able to discover an inductive invariant for an industrial scale dynamic reconfiguration protocol.

In future, our tool can be extended to allow for automatic quantifier template search, and further optimizations can be made to the lemma invariant generation and selection procedures. It would be interesting to explore ways in which the invariant generation procedure can be guided more directly by the generated counterexamples to induction, as a means to prune the search space of candidate invariants more efficiently, perhaps using techniques similar to those presented in \cite{sharma2016invariant}. We would also be interested to see if quantifier structures can be inferred from the protocol syntax itself. Improving the performance of TLC, or experimenting with other, more efficient model checkers \cite{holzmann1997model} would be another avenue, since model checking performance is a main bottleneck of our current approach.

\bibliographystyle{plain}
\bibliography{references.bib}

\begin{thebibliography}{10}

\bibitem{endivetool}
{endive invariant inference tool, Github repository}.
\newblock \url{https://github.com/will62794/endive}, 2022.

\bibitem{mypyvyTool}
mypyvy tool, github repository.
\newblock \url{https://github.com/wilcoxjay/mypyvy}, 2022.

\bibitem{2016abdulla}
Parosh Abdulla, Fr\'{e}d\'{e}ric Haziza, and Luk\'{a}\v{s} Hol\'{\i}k.
\newblock Parameterized verification through view abstraction.
\newblock {\em Int. J. Softw. Tools Technol. Transf.}, 18(5):495–516, Oct
  2016.

\bibitem{arons2001parameterized}
Tamarah Arons, Amir Pnueli, Sitvanit Ruah, Ying Xu, and Lenore Zuck.
\newblock Parameterized verification with automatically computed inductive
  assertions?
\newblock In {\em International Conference on Computer Aided Verification},
  pages 221--234. Springer, 2001.

\bibitem{beersintel08}
Robert Beers.
\newblock {Pre-RTL formal verification: An Intel experience}.
\newblock In {\em 2008 45th ACM/IEEE Design Automation Conference}, pages
  806--811, 2008.

\bibitem{biereBMC}
Armin Biere, Alessandro Cimatti, Edmund Clarke, Ofer Strichman, and Yunshan
  Zhu.
\newblock {Bounded Model Checking}.
\newblock volume~58, pages 117 -- 148, 12 2003.

\bibitem{bloem2015decidability}
Roderick Bloem, Swen Jacobs, Ayrat Khalimov, Igor Konnov, Sasha Rubin, Helmut
  Veith, and Josef Widder.
\newblock {Decidability of Parameterized Verification}.
\newblock {\em Synthesis Lectures on Distributed Computing Theory},
  6(1):1--170, 2015.

\bibitem{2021awsfms3}
James Bornholt, Rajeev Joshi, Vytautas Astrauskas, Brendan Cully, Bernhard
  Kragl, Seth Markle, Kyle Sauri, Drew Schleit, Grant Slatton, Serdar Tasiran,
  Jacob Van~Geffen, and Andrew Warfield.
\newblock {Using Lightweight Formal Methods to Validate a Key-Value Storage
  Node in Amazon S3}.
\newblock In {\em Proceedings of the ACM SIGOPS 28th Symposium on Operating
  Systems Principles}, SOSP '21, page 836–850. Association for Computing
  Machinery, 2021.

\bibitem{2004bouajani}
Ahmed Bouajjani, Peter Habermehl, and Tom{\'a}{\v{s}} Vojnar.
\newblock {Abstract Regular Model Checking}.
\newblock In Rajeev Alur and Doron~A. Peled, editors, {\em Computer Aided
  Verification}, pages 372--386, Berlin, Heidelberg, 2004. Springer Berlin
  Heidelberg.

\bibitem{bradley2011sat}
Aaron~R Bradley.
\newblock {SAT-based model checking without unrolling}.
\newblock In {\em International Workshop on Verification, Model Checking, and
  Abstract Interpretation}, pages 70--87. Springer, 2011.

\bibitem{2020formalspec_tendermint}
Sean Braithwaite, Ethan Buchman, Igor Konnov, Zarko Milosevic, Ilina
  Stoilkovska, Josef Widder, and Anca Zamfir.
\newblock {Formal Specification and Model Checking of the Tendermint Blockchain
  Synchronization Protocol (Short Paper)}.
\newblock In Bruno Bernardo and Diego Marmsoler, editors, {\em 2nd Workshop on
  Formal Methods for Blockchains (FMBC 2020)}, volume~84 of {\em OpenAccess
  Series in Informatics (OASIcs)}, pages 10:1--10:8, Dagstuhl, Germany, 2020.
  Schloss Dagstuhl--Leibniz-Zentrum f{\"u}r Informatik.

\bibitem{burch1992symbolic}
J.~R. Burch, E.~M. Clarke, K.~L. McMillan, D.~L. Dill, and L.~J. Hwang.
\newblock Symbolic model checking: $10^{20}$ states and beyond.
\newblock {\em Inf. Comput.}, 98(2):142–170, jun 1992.

\bibitem{chand2016formal}
Saksham Chand, Yanhong~A Liu, and Scott~D Stoller.
\newblock {Formal Verification of Multi-Paxos for Distributed Consensus}.
\newblock In {\em International Symposium on Formal Methods}, pages 119--136.
  Springer, 2016.

\bibitem{1979Chvatalgreedysetcover}
V.~Chvatal.
\newblock A greedy heuristic for the set-covering problem.
\newblock {\em Math. Oper. Res.}, 4(3):233–235, aug 1979.

\bibitem{clrsthirded}
Thomas~H. Cormen, Charles~E. Leiserson, Ronald~L. Rivest, and Clifford Stein.
\newblock {\em Introduction to Algorithms, 3rd Edition}.
\newblock {MIT} Press, 2009.

\bibitem{cousineau2012tla}
Denis Cousineau, Damien Doligez, Leslie Lamport, Stephan Merz, Daniel Ricketts,
  and Hernan Vanzetto.
\newblock {TLA+ Proofs}.
\newblock {\em Proceedings of the 18th International Symposium on Formal
  Methods (FM 2012), Dimitra Giannakopoulou and Dominique Mery, editors.
  Springer-Verlag Lecture Notes in Computer Science}, 7436:147--154, January
  2012.

\bibitem{de2008z3}
Leonardo De~Moura and Nikolaj Bj{\o}rner.
\newblock {Z3: An efficient SMT solver}.
\newblock In {\em International conference on Tools and Algorithms for the
  Construction and Analysis of Systems}, pages 337--340. Springer, 2008.

\bibitem{2018efficientSATSampling}
Rafael Dutra, Kevin Laeufer, Jonathan Bachrach, and Koushik Sen.
\newblock {Efficient Sampling of SAT Solutions for Testing}.
\newblock In {\em Proceedings of the 40th International Conference on Software
  Engineering}, ICSE '18, page 549–559. Association for Computing Machinery,
  2018.

\bibitem{ernst2007daikon}
Michael~D Ernst, Jeff~H Perkins, Philip~J Guo, Stephen McCamant, Carlos
  Pacheco, Matthew~S Tschantz, and Chen Xiao.
\newblock {The Daikon system for dynamic detection of likely invariants}.
\newblock {\em Science of computer programming}, 69(1-3):35--45, 2007.

\bibitem{2018accelsynsyninvs}
Grigory Fedyukovich and Rastislav Bod{\'i}k.
\newblock {Accelerating Syntax-Guided Invariant Synthesis}.
\newblock In Dirk Beyer and Marieke Huisman, editors, {\em Tools and Algorithms
  for the Construction and Analysis of Systems}, pages 251--269, Cham, 2018.
  Springer International Publishing.

\bibitem{Fedyukovichsyguinvs}
Grigory Fedyukovich, Sumanth Prabhu, Kumar Madhukar, and Aarti Gupta.
\newblock {Quantified Invariants via Syntax-Guided Synthesis}.
\newblock In Isil Dillig and Serdar Tasiran, editors, {\em Computer Aided
  Verification}, pages 259--277, Cham, 2019. Springer International Publishing.

\bibitem{houdiniflanagan}
Cormac Flanagan and K.~Rustan~M. Leino.
\newblock {Houdini, an Annotation Assistant for ESC/Java}.
\newblock In {\em Proceedings of the International Symposium of Formal Methods
  Europe on Formal Methods for Increasing Software Productivity}, FME '01, page
  500–517, Berlin, Heidelberg, 2001. Springer-Verlag.

\bibitem{2010mcmt}
Silvio Ghilardi and Silvio Ranise.
\newblock {MCMT: A Model Checker modulo Theories}.
\newblock In {\em Proceedings of the 5th International Conference on Automated
  Reasoning}, IJCAR'10, page 22–29, Berlin, Heidelberg, 2010.
  Springer-Verlag.

\bibitem{goel2021symmetry}
Aman Goel and Karem Sakallah.
\newblock {On Symmetry and Quantification: A New Approach to Verify Distributed
  Protocols}.
\newblock In {\em NASA Formal Methods Symposium}, pages 131--150. Springer,
  2021.

\bibitem{hance2021finding}
Travis Hance, Marijn Heule, Ruben Martins, and Bryan Parno.
\newblock {Finding Invariants of Distributed Systems: It{\textquoteright}s a
  Small (Enough) World After All}.
\newblock In {\em 18th USENIX Symposium on Networked Systems Design and
  Implementation (NSDI 21)}, pages 115--131. USENIX Association, April 2021.

\bibitem{holzmann1997model}
Gerard~J. Holzmann.
\newblock {The model checker SPIN}.
\newblock {\em IEEE Transactions on software engineering}, 23(5):279--295,
  1997.

\bibitem{softwareabstractions}
Daniel Jackson.
\newblock {\em Software Abstractions - Logic, Language, and Analysis}.
\newblock MIT Press, 2006.

\bibitem{Karp1972}
Richard~M. Karp.
\newblock {\em Reducibility among Combinatorial Problems}, pages 85--103.
\newblock Springer US, Boston, MA, 1972.

\bibitem{2020firstorderquantified}
Jason~R. Koenig, Oded Padon, Neil Immerman, and Alex Aiken.
\newblock {First-Order Quantified Separators}.
\newblock In {\em Proceedings of the 41st ACM SIGPLAN Conference on Programming
  Language Design and Implementation}, PLDI 2020, page 703–717. Association
  for Computing Machinery, 2020.

\bibitem{2019tlamadesymbolic}
Igor Konnov, Jure Kukovec, and Thanh-Hai Tran.
\newblock {TLA+ Model Checking Made Symbolic}.
\newblock {\em Proc. ACM Program. Lang.}, 3(OOPSLA), Oct 2019.

\bibitem{kuppe2017verified}
Markus~A Kuppe.
\newblock {A Verified and Scalable Hash Table for the TLC Model Checker:
  Towards an Order of Magnitude Speedup}.
\newblock Master's thesis, University of Hamburg., 2017.

\bibitem{lamport1998the}
Leslie Lamport.
\newblock {The Part-Time Parliament}.
\newblock {\em ACM Trans. Comput. Syst.}, 16(2):133–169, May 1998.

\bibitem{lamport2002specifying}
Leslie Lamport.
\newblock {\em {Specifying Systems: The TLA+ Language and Tools for Hardware
  and Software Engineers}}.
\newblock Addison-Wesley, Jun 2002.

\bibitem{Lamport2018UsingTT}
Leslie Lamport.
\newblock {Using TLC to Check Inductive Invariance}.
\newblock \url{http://lamport.azurewebsites.net/tla/inductive-invariant.pdf},
  2018.

\bibitem{Ma2019}
Haojun Ma, Aman Goel, Jean~Baptiste Jeannin, Manos Kapritsos, Baris Kasikci,
  and Karem~A. Sakallah.
\newblock {I4: Incremental Inference of Inductive Invariants for Verification
  of Distributed Protocols}.
\newblock In {\em SOSP 2019 - Proceedings of the 27th ACM Symposium on
  Operating Systems Principles}, 2019.

\bibitem{mannasafetybook}
Zohar Manna and Amir Pnueli.
\newblock {\em Temporal Verification of Reactive Systems: Safety}.
\newblock Springer-Verlag, Berlin, Heidelberg, 1995.

\bibitem{2016merzmanysorted}
Stephan Merz and Hern{\'a}n Vanzetto.
\newblock {Encoding TLA+ into Many-Sorted First-Order Logic}.
\newblock In Michael Butler, Klaus-Dieter Schewe, Atif Mashkoor, and Miklos
  Biro, editors, {\em Abstract State Machines, Alloy, B, TLA, VDM, and Z},
  pages 54--69, Cham, 2016. Springer International Publishing.

\bibitem{newcombe2014use}
Chris Newcombe, Tim Rath, Fan Zhang, Bogdan Munteanu, Marc Brooker, and Michael
  Deardeuff.
\newblock {How Amazon Web Services Uses Formal Methods}.
\newblock {\em Commun. ACM}, 58(4):66–73, March 2015.

\bibitem{raftpaper}
Diego Ongaro and John Ousterhout.
\newblock {In Search of an Understandable Consensus Algorithm}.
\newblock In {\em Proceedings of the 2014 USENIX Conference on USENIX Annual
  Technical Conference}, USENIX ATC'14, pages 305--320, USA, 2014. USENIX
  Association.

\bibitem{padon2016decidability}
Oded Padon, Neil Immerman, Sharon Shoham, Aleksandr Karbyshev, and Mooly Sagiv.
\newblock {Decidability of Inferring Inductive Invariants}.
\newblock In Rastislav Bod{\'{\i}}k and Rupak Majumdar, editors, {\em
  Proceedings of the 43rd Annual {ACM} {SIGPLAN-SIGACT} Symposium on Principles
  of Programming Languages, {POPL} 2016, St. Petersburg, FL, USA, January 20 -
  22, 2016}, pages 217--231. {ACM}, 2016.

\bibitem{padonpaxosEPR}
Oded Padon, Giuliano Losa, Mooly Sagiv, and Sharon Shoham.
\newblock {Paxos Made EPR: Decidable Reasoning about Distributed Protocols}.
\newblock {\em Proc. ACM Program. Lang.}, 1(OOPSLA), Oct 2017.

\bibitem{padon2016ivy}
Oded Padon, Kenneth~L McMillan, Aurojit Panda, Mooly Sagiv, and Sharon Shoham.
\newblock {Ivy: Safety Verification by Interactive Generalization}.
\newblock In {\em Proceedings of the 37th ACM SIGPLAN Conference on Programming
  Language Design and Implementation}, pages 614--630, 2016.

\bibitem{fmcad22artifact}
William Schultz, Ian Dardik, and Stavros Tripakis.
\newblock {Artifact for FMCAD 2022 paper: Plain and Simple Inductive Invariant
  Inference for Distributed Protocols in TLA+}.
\newblock \url{https://doi.org/10.5281/zenodo.6994922}, August 2022.

\bibitem{schultz2021formal}
William Schultz, Ian Dardik, and Stavros Tripakis.
\newblock {Formal Verification of a Distributed Dynamic Reconfiguration
  Protocol}.
\newblock In {\em Proceedings of the 11th ACM SIGPLAN International Conference
  on Certified Programs and Proofs}, CPP 2022, page 143–152, Philadelphia,
  PA, USA, 2022. Association for Computing Machinery.

\bibitem{schultz2021design}
William Schultz, Siyuan Zhou, Ian Dardik, and Stavros Tripakis.
\newblock {Design and Analysis of a Logless Dynamic Reconfiguration Protocol}.
\newblock In Quentin Bramas, Vincent Gramoli, and Alessia Milani, editors, {\em
  25th International Conference on Principles of Distributed Systems (OPODIS
  2021)}, volume 217 of {\em Leibniz International Proceedings in Informatics
  (LIPIcs)}, pages 26:1--26:16, Dagstuhl, Germany, 2022. Schloss Dagstuhl --
  Leibniz-Zentrum f{\"u}r Informatik.

\bibitem{sharma2016invariant}
Rahul Sharma and Alex Aiken.
\newblock {From Invariant Checking to Invariant Inference Using Randomized
  Search}.
\newblock In Armin Biere and Roderick Bloem, editors, {\em Computer Aided
  Verification - 26th International Conference, {CAV} 2014, Held as Part of the
  Vienna Summer of Logic, {VSL} 2014, Vienna, Austria, July 18-22, 2014.
  Proceedings}, volume 8559 of {\em Lecture Notes in Computer Science}, pages
  88--105. Springer, 2014.

\bibitem{2020cockroachdb}
Rebecca Taft, Irfan Sharif, Andrei Matei, Nathan VanBenschoten, Jordan Lewis,
  Tobias Grieger, Kai Niemi, Andy Woods, Anne Birzin, Raphael Poss, Paul
  Bardea, Amruta Ranade, Ben Darnell, Bram Gruneir, Justin Jaffray, Lucy Zhang,
  and Peter Mattis.
\newblock {CockroachDB: The Resilient Geo-Distributed SQL Database}.
\newblock In {\em Proceedings of the 2020 ACM SIGMOD International Conference
  on Management of Data}, SIGMOD '20, page 1493–1509. Association for
  Computing Machinery, 2020.

\bibitem{Wernick1942CompleteSO}
William Wernick.
\newblock Complete sets of logical functions.
\newblock {\em Transactions of the American Mathematical Society}, 51:117--132,
  1942.

\bibitem{wilcox2015verdi}
James~R. Wilcox, Doug Woos, Pavel Panchekha, Zachary Tatlock, Xi~Wang,
  Michael~D. Ernst, and Thomas~E. Anderson.
\newblock Verdi: a framework for implementing and formally verifying
  distributed systems.
\newblock In David Grove and Stephen~M. Blackburn, editors, {\em Proceedings of
  the 36th {ACM} {SIGPLAN} Conference on Programming Language Design and
  Implementation, Portland, OR, USA, June 15-17, 2015}, pages 357--368. {ACM},
  2015.

\bibitem{woos2016planning}
Doug Woos, James~R. Wilcox, Steve Anton, Zachary Tatlock, Michael~D. Ernst, and
  Thomas Anderson.
\newblock {Planning for Change in a Formal Verification of the Raft Consensus
  Protocol}.
\newblock In {\em Proceedings of the 5th ACM SIGPLAN Conference on Certified
  Programs and Proofs}, CPP 2016, page 154–165. Association for Computing
  Machinery, 2016.

\bibitem{yao2021distai}
Jianan Yao, Runzhou Tao, Ronghui Gu, Jason Nieh, Suman Jana, and Gabriel Ryan.
\newblock {{DistAI}: {Data-Driven} Automated Invariant Learning for Distributed
  Protocols}.
\newblock In {\em 15th USENIX Symposium on Operating Systems Design and
  Implementation (OSDI 21)}, pages 405--421. USENIX Association, July 2021.

\bibitem{tlcmodelchecker}
Yuan Yu, Panagiotis Manolios, and Leslie Lamport.
\newblock {Model Checking TLA+ Specifications}.
\newblock In Laurence Pierre and Thomas Kropf, editors, {\em Correct Hardware
  Design and Verification Methods}, pages 54--66, Berlin, Heidelberg, 1999.
  Springer Berlin Heidelberg.

\end{thebibliography}

\appendices

\section{Detailed Benchmark Results}
\label{sec:appendix-results}

Table~\ref{fig:benchmark-results-detailed} gives a more detailed breakdown of the results presented in Table~\ref{fig:benchmark-results} for our \toolendive{} invariant inference tool. The \textit{Check}, \textit{Elim}, and \textit{CTIGen} columns of Table~\ref{fig:benchmark-results-detailed} indicate, respectively, the wall clock time in seconds for (1)  checking candidate lemma invariants, (2) eliminating CTIs, and (3) generating CTIs. The \textit{CTIs} column indicates the total number of eliminated CTIs. 

Recall that we limit the maximum number of generated CTIs to 10000 per round, as mentioned in Section \ref{sec:impl-and-experiment-setup}. This explains why some protocol results for the \toolendive{} tool report elimination of exactly 10000 CTIs. For example, for the tla-twophase benchmark, an inductive invariant was discovered in a single round of the algorithm loop (starting at Line \ref{algolineloopbegin} of Algorithm \ref{fig:invgen-algo}), so no more than 10000 CTIs were generated in the entire run. If the benchmark run eliminated greater than 10000 CTIs, this indicates that it ran for more more than 1 round.

Also, for protocols that eliminated 0 CTIs (e.g. tla-consensus, tla-tcommit), this indicates that the starting safety property was already inductive. Thus, no CTIs were ever generated and no lemma invariants were needed. Similarly, some protocols eliminated a nonzero amount of CTIs less than 10000 (e.g. ex-quorum-leader-election). This may be the case when no more than a single round of the algorithm was needed to discover an inductive invariant, or that the number of generated counterexamples at each round did not exceed 10000.  Recall that, even within a single round of the algorithm, as shown in Algorithm \ref{fig:invgen-algo}, it is possible to discover multiple new lemma invariants.

Additional statistics on the instance sizes used during invariant inference and the degree of automation required for TLAPS proofs are shown in Table~\ref{fig:benchmark-results-detailed-extra}. The \textit{TLAPS Auto} column indicates whether the TLAPS proof of the inductive invariant discovered by \toolendive{} was completely automatic (indicated with a \checkmark), or required some user assistance (indicated with a \xmark). 

To provide more fine-grained detail on the level of automation for each TLAPS proof, the \textit{TLAPS Auto} column also includes the number of verification conditions in the induction check that were proved fully automatically. For a protocol with a transition relation of the form $Next = T_1 \vee \dots \vee T_k$ and an inductive invariant candidate $Ind = A_1 \wedge \dots \wedge A_n$, the consecution check $Ind \wedge Next \Rightarrow Ind'$ is typically the most significant verification burden, and can be trivially decomposed into $k \cdot n$ verification conditions (VCs). That is, a verification condition $Ind \wedge T_j \Rightarrow A_i'$ is generated for each $j \in \{1,\dots, k\}$ and $i \in \{1,\dots, n\}$, giving $k \cdot n$ total VCs. We notate these statistics in the \textit{TLAPS Auto} column as (\text{\# VCs proved automatically} / \text{$k \cdot n$ total VCs}). Protocols that were proved fully automatically are shown as $(k \cdot n/k \cdot n)$. The \textit{Check (s)} column also shows the total time in seconds needed to check each proof, as measured on a 2020 M1 Macbook Air using version 1.4.5 of the TLA+ proof manager.

\begin{table}[h]
    \caption{Detailed profiling results for the \toolendive{} results from Table \ref{fig:benchmark-results}.}
    \label{fig:benchmark-results-detailed}
    \begin{center}
        \begingroup
        \setlength{\tabcolsep}{2pt} 
        \renewcommand{\arraystretch}{1.0} 
        \scriptsize
        \begin{tabular}{ l | l | r r r r r }
\hline
No. & Protocol & Time & CTIs & Check & Elim & CTIGen\\ \hline 
1 & tla-consensus & 1 & 0 & 0 & 0 & 1 \\ 
2 & tla-tcommit & 2 & 0 & 0 & 0 & 2 \\ 
3 & i4-lock-server & 7 & 12 & 2 & 2 & 4 \\ 
4 & ex-quorum-leader-election & 11 & 204 & 2 & 2 & 7 \\ 
5 & pyv-toy-consensus-forall & 19 & 412 & 2 & 2 & 15 \\ 
6 & tla-simple & 8 & 15 & 2 & 2 & 5 \\ 
7 & ex-lockserv-automaton & 23 & 3624 & 6 & 8 & 9 \\ 
8 & tla-simpleregular & 10 & 1972 & 3 & 3 & 5 \\ 
9 & pyv-sharded-kv & 312 & 11715 & 17 & 46 & 249 \\ 
10 & pyv-lockserv & 35 & 3654 & 11 & 11 & 13 \\ 
11 & tla-twophase & 43 & 10000 & 10 & 22 & 12 \\ 
12 & i4-learning-switch & \gray{TO} &  &  &  \\
13 & ex-simple-decentralized-lock & 44 & 2035 & 13 & 18 & 14 \\ 
14 & i4-two-phase-commit & 69 & 10408 & 18 & 19 & 33 \\ 
15 & pyv-consensus-wo-decide & 127 & 12995 & 56 & 39 & 32 \\ 
16 & pyv-consensus-forall & 175 & 10609 & 63 & 25 & 88 \\ 
17 & pyv-learning-switch & \gray{TO} &  &  &  \\
18 & i4-chord-ring-maintenance & \gray{n/a} &  &  &  \\
19 & pyv-sharded-kv-no-lost-keys & 13 & 404 & 2 & 2 & 9 \\ 
20 & ex-naive-consensus & 40 & 10000 & 10 & 15 & 16 \\ 
21 & pyv-client-server-ae & 46 & 10000 & 2 & 4 & 40 \\ 
22 & ex-simple-election & 24 & 551 & 10 & 7 & 8 \\ 
23 & pyv-toy-consensus-epr & 19 & 384 & 8 & 6 & 6 \\ 
24 & ex-toy-consensus & 7 & 14 & 2 & 2 & 4 \\ 
25 & pyv-client-server-db-ae & 4941 & 12546 & 4657 & 46 & 239 \\ 
26 & pyv-hybrid-reliable-broadcast & \gray{n/a} &  &  &  \\
27 & pyv-firewall & 38 & 1740 & 11 & 22 & 7 \\ 
28 & ex-majorityset-leader-election & 53 & 10000 & 12 & 15 & 26 \\ 
29 & pyv-consensus-epr & 247 & 16269 & 80 & 38 & 129 \\ 
\hline
30 & mldr & 2025 & 7751 & 1272 & 651 & 102 \\ 
\end{tabular}

        \endgroup
    \end{center}
\end{table}

\begin{table}[h]
    \caption{Additional statistics for \toolendive{} results reported in Table \ref{fig:benchmark-results}.}
    \label{fig:benchmark-results-detailed-extra}
    \begin{center}
        \begingroup
        \setlength{\tabcolsep}{2pt} 
        \renewcommand{\arraystretch}{1.0} 
        \scriptsize
        \newcommand{\autoyes}{\checkmark\,}
\newcommand{\autono}{\xmark\,}

\begin{tabular}{| l | l | p{2.2cm} | c | c | }
\hline
No. & Protocol & Instance Size & TLAPS Auto & Check (s) \\ \hline 
1 & tla-consensus & \text{Value=\{v1,v2,v3\}} & \autoyes (1/1) & 13 \\  \hline 
2 & tla-tcommit & \text{RM=\{rm1,rm2,rm3\}} & \autoyes (2/2) & 1 \\ \hline  
3 & i4-lock-server & \text{Server=\{s1,s2\}} \text{Client=\{c1,c2\}} & \autoyes (4/4) & 1\\ \hline 
4 & ex-quorum-leader-election & \text{Node=\{n1,n2,n3,n4\}} & \autoyes (4/4) &  1 \\ \hline
5 & pyv-toy-consensus-forall & \text{Node=\{n1,n2,n3\}} \text{Value=\{v1,v2\}} & \autoyes (6/6) & 1  \\ \hline 
6 & tla-simple & \text{N=4} & \autoyes (4/4) & 1 \\ \hline 
7 & ex-lockserv-automaton & \text{Node=\{n1,n2,n3\}} & \autoyes (45/45) & 6 \\ \hline
8 & tla-simpleregular & \text{N=3} & \autoyes (12/12) &  1 \\ \hline  
9 & pyv-sharded-kv & \text{Node=\{n1,n2,n3\}} \text{Key=\{k1,k2\}} \text{Value=\{v1,v2\}} & \autoyes (18/18) & 15  \\ \hline 
10 & pyv-lockserv & \text{Node=\{n1,n2,n3\}} & \autoyes (45/45) & 6 \\ \hline
11 & tla-twophase & \text{RM=\{rm1,rm2,rm3\}} & \autono (68/70) & 18 \\ \hline 
12 & i4-learning-switch & \gray{TO} & & \\\hline 
13 & ex-simple-decentralized-lock & \text{Node=\{n1,n2,n3\}} & \autoyes (8/8) & 17  \\ \hline
14 & i4-two-phase-commit & \text{Node=\{n1,n2,n3\}} & \autoyes (77/77) & 6  \\ \hline 
15 & pyv-consensus-wo-decide & \text{Node=\{n1,n2,n3\}} & \autono (35/40) & 20  \\ \hline 
16 & pyv-consensus-forall & \text{Node=\{n1,n2,n3\}} & \autono (46/48) & 25  \\ \hline 
17 & pyv-learning-switch & \gray{TO} & & \\ \hline 
18 & i4-chord-ring-maintenance & \gray{n/a} & &  \\ \hline 
19 & pyv-sharded-kv-no-lost-keys & \text{Node=\{n1,n2\}} \text{Key=\{k1,k2\}} \text{Value=\{v1,v2\}} & \autoyes (6/6) & 12  \\ \hline
20 & ex-naive-consensus & \text{Node=\{n1,n2,n3\}}  \text{Value=\{v1,v2\}} & \autono (11/12)  & 6  \\ \hline
21 & pyv-client-server-ae & \text{Node=\{n1,n2,n3\}} \text{Request=\{r1,r2\}} \text{Response=\{p1,p2\}} & \autoyes (6/6)  & 2  \\ \hline
22 & ex-simple-election & \text{Acceptor=\{a1,a2,a3\}} \text{Proposer=\{p1,p2\}} & \autono (11/12)  & 5  \\ \hline
23 & pyv-toy-consensus-epr & \text{Node=\{n1,n2,n3\}} \text{Value=\{v1,v2\}} & \autono (6/8)  & 9  \\ \hline
24 & ex-toy-consensus & \text{Node=\{n1,n2,n3\}} \text{Value=\{v1,v2\}} & \autono (1/4)  & 1  \\ \hline
25 & pyv-client-server-db-ae & \text{Node=\{n1,n2,n3\}} \text{Request = \{r1,r2,r3\}} \text{Response=\{p1,p2,p3\}} \text{DbRequestId=\{i1,i2\}} & \autoyes (40/40)  & 20  \\ \hline
26 & pyv-hybrid-reliable-broadcast & \gray{n/a} & & \\ \hline 
27 & pyv-firewall & \text{Node=\{n1,n2,n3\}} & \autono (4/10)  & 23 \\  \hline 
28 & ex-majorityset-leader-election & \text{Node=\{n1,n2,n3\}} & \autono (9/12)  & 9 \\ \hline 
29 & pyv-consensus-epr & \text{Node=\{n1,n2,n3\}} \text{Value=\{v1,v2\}} & \autono (39/40)  & 21 \\ 
\hline
30 & mldr & \text{MaxTerm=3} \text{MaxConfigVersion=3} \text{Server=\{n1,n2,n3,n4\}} & \autono (15/24)  & 226 \\ \hline  
\end{tabular}

        \endgroup
    \end{center}
\end{table}

\end{document}